\documentclass[lettersize,journal]{IEEEtran}
\usepackage{amsmath,amsfonts}
\usepackage{algorithmic}
\usepackage{algorithm}
\usepackage{array}
\usepackage[caption=false,font=normalsize,labelfont=sf,textfont=sf]{subfig}
\usepackage{textcomp}
\usepackage{stfloats}
\usepackage{url}
\usepackage{verbatim}
\usepackage{graphicx}
\usepackage{cite}
\usepackage{multirow,multicol,tabularx}
\usepackage{amsmath, amssymb, amsthm}
\theoremstyle{plain}

\newtheorem{theorem}{Theorem}
\newtheorem{lemma}{Lemma}

\begin{document}

\title{Self-supervised training of deep denoisers in multi-coil MRI considering noise correlations}
 
\author{Juhyung Park, Dongwon Park, Sooyeon Ji, Hyeong-Geol Shin, \\ Se Young Chun, \IEEEmembership{Member, IEEE}, and Jongho Lee, \IEEEmembership{Member, IEEE}
\thanks{J. Park is the first author and S.Y. Chun and J. Lee are corresponding authors.
This paper was produced by the IEEE Publication Technology Group. They are in Piscataway, NJ.}
\thanks{Manuscript received xxx xx, xxxx; revised xxx xx, xxxx.}
\thanks{J. Park, S. Y. Chun and J. Lee are with the Department of Electrical and Computer Engineering, Seoul National University, Republic of Korea. (e-mail: jack0878@snu.ac.kr; sychun@snu.ac.kr; jonghoyi@snu.ac.kr).}%
\thanks{D. Park is with INMC \& IPAI, Seoul National University, Republic of Korea. (email: dong1park@snu.ac.kr).}
\thanks{S. Ji is with the Division of Computer Engineering, Hankuk University of Foreign Studies, Yongin, South Korea. (e-mail: sueji0221@hufs.ac.kr).}
\thanks{H-G. Shin is with the Department of Radiology, Johns Hopkins University School of Medicine, Baltimore, MD and F.M. Kirby Research Center for Functional Brain Imaging, Kennedy Krieger Institute, Baltimore, MD (e-mail: sin4109@gmail.com).}
}
\markboth{IEEE Transactions on Computational Imaging}%
{Park \MakeLowercase{\textit{et al.}}: Coil2Coil}

\maketitle
 
\begin{abstract}
Deep learning-based denoising methods have shown powerful results for improving the signal-to-noise ratio of magnetic resonance (MR) images, mostly by leveraging supervised learning with clean ground truth. However, acquiring clean ground truth images is often expensive and time-consuming. Self-supervised methods have been widely investigated to mitigate the dependency on clean images, but mostly rely on the suboptimal splitting of K-space measurements of an image to yield input and target images for ensuring statistical independence. In this study, we investigate an alternative self-supervised training method for deep denoisers in multi-coil MRI, dubbed Coil2Coil (C2C), that naturally split and combine the multi-coil data among phased-array coils, generating two noise-corrupted images for training. This novel approach allows exploiting multi-coil redundancy, but the images are statistically correlated and may not have the same clean image. To mitigate these issues, we propose the methods to approximately decorrelate the statistical dependence of these images and match the underlying clean images, thus enabling them to be used as the training pairs. For synthetic denoising experiments, C2C yielded the best performance (pSNR: 39.01 ± 2.91) against prior self-supervised methods, reporting outcome comparable even to supervised methods. For real-world denoising cases, C2C yielded consistent performance as synthetic cases, removing only noise structures.
\end{abstract}


\section{Introduction}
\IEEEPARstart{M}{agnetic} resonance imaging (MRI) is a powerful modality for in-vivo imaging. MR images often suffer from signal-to-noise ratio (SNR) deficiency~\cite{ref1}, hampering accurate diagnosis or proper processing. Therefore, enhancing image SNR has long been an important research topic in MR image processing. One popular approach for improving SNR is using denoising algorithms~\cite{ref2}, which are designed as post-processing tools and can be applied to existing images. Traditionally, algorithms such as filter-based methods~\cite{ref3,ref4,ref5} and low-rank-based methods~\cite{ref6,ref7} have been developed.

Recently, deep learning-based denoising methods have been widely investigated~\cite{ref8,ref9}. In MRI, deep learning has been successfully applied in various processing steps~\cite{ref10,ref11,ref12,ref13}, including denoising~\cite{ref14,ref15,ref16,ref17,ref18,ref19,ref20,ref21}, demonstrating that it outperforms the conventional denoising methods. These deep learning based MRI denoising methods were largely developed based on supervised learning. In the supervised learning, the network is trained on pairs of images, where each pair consist of a noise-corrupted images as an input and the corresponding clean image as an label. aimed to improve the SNR of the noise-corrupted image~\cite{ref22}.

In most cases, however, obtaining a clean image is expensive or practically infeasible. To address this limitation, Noise2Noise (N2N)~\cite{ref23} was proposed, which can train a network without clean images. This method uses pairs of noise-corrupted images that have a same clean image but are corrupted by uncorrelated noise realizations, relaxing the requirement for clean images in the supervised learning. Despite this advantage, N2N still has limited applicability due to the necessity of having paired noise-corrupted images~\cite{ref24}. This challenges is especially pronounced in MRI due to motion artifacts or filed inhomogeneity between two acquisitions.
 
Subsequently, various self-supervised learning methods have been proposed to overcome this limitation by using only a single noise-corrupted image ~\cite{ref25,ref26,ref27,ref28,ref29,ref30,ref31,ref32,ref33,ref34,ref35}. While these methods mitigate the need for the clean image or the noise-corrupted pairs, they often suffer from degraded denoising performance~\cite{ref25,ref26,ref27,ref28,ref31,ref32,ref33,ref34,ref35}, or require an accurate prior knowledge of the noise model and its level~\cite{ref29, ref30}.

In MR image denoising, various methods which do not require clean image have been proposed~\cite{ref23, ref36, ref37, ref38}. For example, a two noise-corrupted images generation scheme within a single image was proposed using compressed-sensing (CS)~\cite{ref38} by splitting K-space measurements into two to ensure the statistical independence of them, which enables the use of N2N training~\cite{ref23}. However, since the two CS-reconstructed images do not perfectly share the same clean image, generating a unnatural image far from the real MR image resulted in performance degradation. Other denoising methods have been proposed by utilizing redundancy in a specific MR sequence (\textit{e.g.}, redundancy of b-value domain in diffusion-weighted imaging (DWI)~\cite{ref36, ref37} and temporal domain in dynamic MR imaging~\cite{ref39}). Although these methods are free from the constraint of obtaining clean images, the usability of these methods is limited to specific MR acquisitions (\textit{i.e.}, DWI and dynamic MR).
		
In this study, we propose a novel alternative self-supervised training method, referred to as Coil2Coil (C2C), that generates a pair of noise-corrupted images using multi-coil images from phased array coils in order to exploit multi-coil redundancy. The multi-coil images were split into two groups and combined, further processed to share the same clean image while having uncorrelated noise, thereby generating an independent noise-corrupted pair for N2N training.
We evaluated the performance of our proposed method on denoising experiments that were performed with synthetic noise-added images as well as real-world DICOM images, demonstrating excellent performance in multi-coil MRI denoising.
Note that the source code of C2C is available on GitHub\footnote{https://github.com/SNU-LIST/Coil2Coil}.

\section{Related Works}

\subsection{Noise2Noise}

In N2N, a denoising network is trained with the pairs of noise-corrupted images, with one image as an input and the other image as a label, learning the characteristics of noise between the two images for denoising~\cite{ref23}. The method requires the following three conditions: (1) The paired images have uncorrelated noise. (2) They have the same clean image. (3) The expectation of the noise is zero. When training a denoising network, an L2 loss between the output and label is used because this L2 loss is equal to that between the output and clean image. For MRI denoising, a paired noise-corrupted images generation scheme was proposed based on CS~\cite{ref39}, where the K-space signal of single noise-corrupted image is split into two subset and each is reconstructed using CS. While this pair demonstrates the potential of N2N framework, the resulting CS-reconstructed images do not perfectly share clean signal, making them unnatural and far from the true MR image. 

\subsection{Self-supervised denoising methods}

To overcome the limitation of acquiring a clean image or noise-corrupted pairs, Further studies have been proposed by utilizing only a single noise-corrupted image for training~\cite{ref25,ref26,ref27,ref28,ref29,ref30,ref31,ref32,ref33,ref34,ref35}. Various studies proposed the masking strategies, which masking out some pixels in the input image and utilizing the masked pixels as a label~\cite{ref25,ref26,ref27}. Depending on a mask generation strategy, the method is referred to as Noise2Void (N2V: when the masked pixels are randomly chosen~\cite{ref25}), Noise2Self (N2Se: when a masking scheme is suggested based on the independence of the input image and label pixels~\cite{ref26}), or Noise2Same (N2Sa: when a self-similarity loss between input and denoised images is utilized~\cite{ref27}). Neighbor2Neighbor (Ne2Ne)~\cite{ref28} proposed a sub-sampling method which generates multiple down-sampled image pairs within a single image. However, the resulting images may contain correlated noise and have different distribution from target domain. In addition, several studies has adopted Stein’s unbiased risk estimator for deonising~\cite{ref29, ref30}, although these methods require accurate prior knowledge of the noise model and its level. More recent studies have been proposed~\cite{ref31, ref32, ref33, ref34, ref35}, which are specialized for the image domain and do not sufficiently consider MR imaging, limiting their ability to produce optimal results in the MR domain.

\begin{figure*}[b]
\centering
\includegraphics[width=0.85\textwidth]{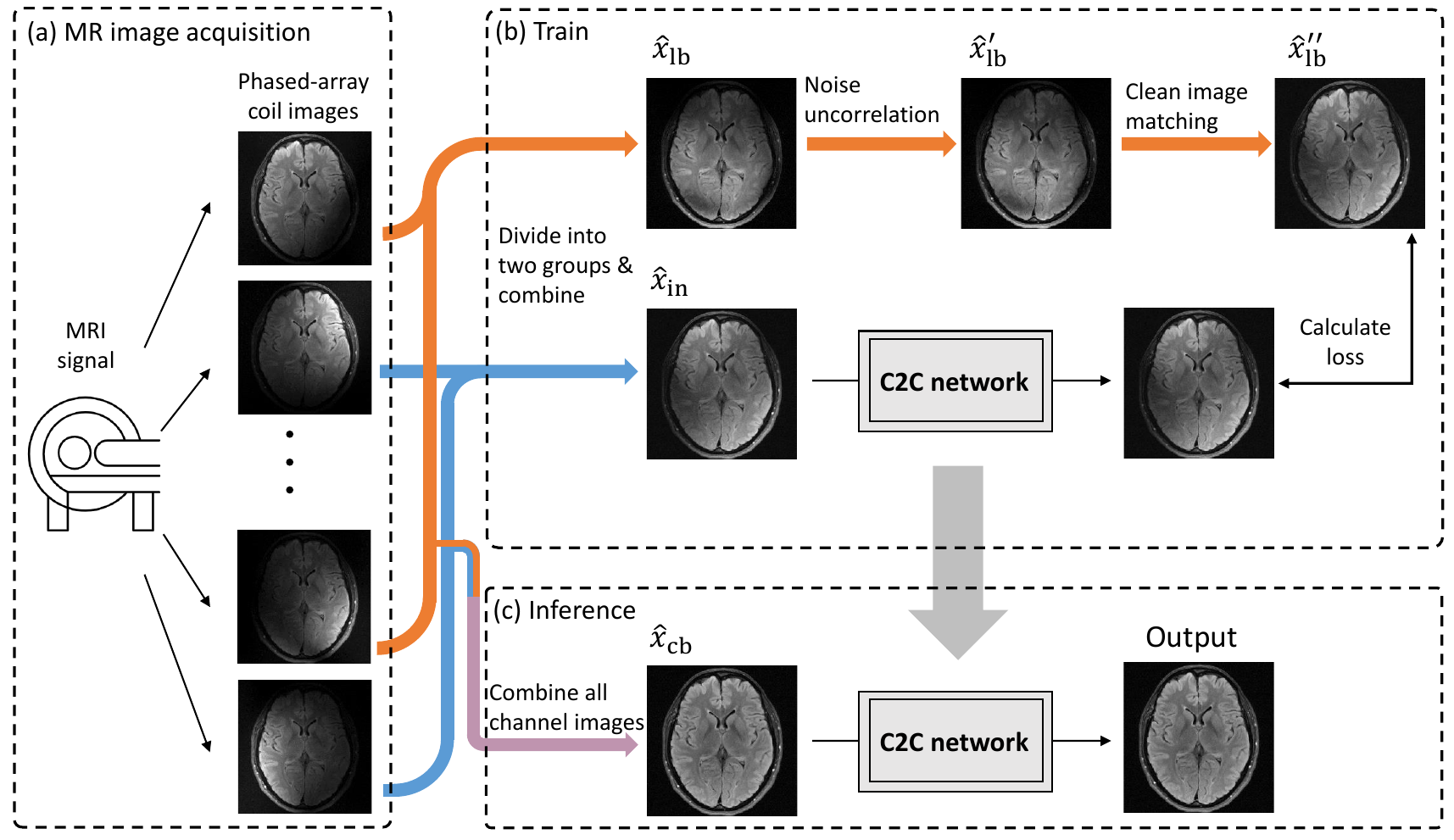}
\caption{Overview of Coil2Coil (C2C). (a) When using a phased-array coil, multiple coil images are acquired. 
(b) In C2C, these multiple coil images are divided into two groups and combined to generate $\hat{\mathbf{x}}_{\text{in}}$ and $\hat{\mathbf{x}}_{\text{lb}}$. Then, $\hat{\mathbf{x}}_{\text{lb}}$ is processed to have uncorrelated noise and same clean image as $\hat{\mathbf{x}}_{\text{in}}$, to use the N2N training scheme. 
Finally, this processed image ($\hat{\mathbf{x}}^{''}_{\text{in}}$) paired with $\hat{\mathbf{x}}_{\text{in}}$ are utilized for the training of a denoising neural network. 
(c) For inference, all channel images are combined (e.g., DICOM image) and used as an input for the network, creating a denoised image.}
\label{fig_1}
\end{figure*}

\section{Methods}

\subsection{MRI acquisition}

When using a phased array coil to acquire an MR image $\mathbf{x} \in \mathbb{R}^m$, an individual coil image  \(\mathbf{y}^{(i)} \in \mathbb{C}^m\) of the $i$th coil $i \in \{1, \ldots, n\}$ can be formulated as 
\begin{eqnarray}
\mathbf{y}^{(i)} = \mathbf{s}^{(i)} \cdot \mathbf{x} + \mathbf{n}_{c}^{(i)}
\end{eqnarray}
where $\cdot$ denoted the Hadamard product, \(\mathbf{s}^{(i)} \in \mathbb{C}^m\) representing coil sensitivity and \(\mathbf{n}_{c}^{(i)} = \mathbf{n}_{r}^{(i)} + i \mathbf{n}_{i}^{(i)}\) is the complex noise in the \(i\)th coil, modeled as zero-mean Gaussian noise with a standard deviation of \(\sigma^{(i)}\) for both real and imaginary components: $\mathbf{n}_{r}^{(i)} \in \mathbb{R}^m$, and $\mathbf{n}_{i}^{(i)} \in \mathbb{R}^m$.  

A combined MR image (\(\hat{\mathbf{x}}_{\text{cb}} \in \mathbb{R}^m\)) using multi-coil images can be formulated as follows~\cite{ref40}: 
\begin{eqnarray}
\begin{split}
\hat{\mathbf{x}}_{\text{cb}} &=\left| \sum_i \mathbf{s}^{(i)H} \cdot \mathbf{y}  \right|
\end{split}
\end{eqnarray}
Where $\mathbf{s}^{(i)H}$ is hermitian of $\mathbf{s}^{(i)}$.

\subsection{Revisiting Noise2Noise}

The N2N method has been proposed to train deep neural networks only with noise-corrupted images where two noise realizations per image were required~\cite{ref23}. In this work, its theoretical justification required zero-mean noise, but there was no clear assumption on independence or uncorrelated property of two realizations. Zhussip et al. investigated this briefly in \cite{ref29} and the following lemma was proposed for two independent noise realizations per image.
\begin{lemma}[Zhussip et al.~\cite{ref29}]
Suppose that the triplet $(\mathbf{x}, \mathbf{y}, \mathbf{z})$ follows a joint distribution. If two noise vectors $\mathbf{y} - \mathbf{x}$, $\mathbf{z} - \mathbf{x}$ are both zero mean, and $\mathbf{y}$ is uncorrelated or independent of ($\mathbf{z} - \mathbf{x}$), the following equation holds:
\begin{eqnarray}
\mathbb{E}_{(\mathbf{x}, \mathbf{y})}
\| \mathbf{x} - \boldsymbol{h_\theta}( \mathbf{y} ) \|^2 
= \mathbb{E}_{(\mathbf{x}, \mathbf{y}, \mathbf{z})} \left \| \mathbf{z} - \boldsymbol{h_\theta}( \mathbf{y} ) \|^2 \right..
\end{eqnarray}
\end{lemma}
\begin{proof} The mean-squared error (MSE) can be rewritten as follows:
\begin{eqnarray}
&& \mathbb{E}_{(\mathbf{x}, \mathbf{y})}
\| \mathbf{x} - \boldsymbol{h_\theta}( \mathbf{y} ) \|^2  \\ 
&=&
\mathbb{E}_{\mathbf{x}}  \mathbb{E}_{(\mathbf{y}, \mathbf{z}) | \mathbf{x} } 
\left\{ \| \mathbf{x} - \mathbf{z} + \mathbf{z} - \boldsymbol{h_\theta}( \mathbf{y} ) \|^2 | \mathbf{x} \right\} \nonumber \\
&=& \mathbb{E}_{\mathbf{x}} \mathbb{E}_{(\mathbf{y}, \mathbf{z}) | \mathbf{x}} 
\left\{ \| \mathbf{z} - \boldsymbol{h_\theta}( \mathbf{y} ) \|^2 + 2  (\mathbf{z} - \mathbf{x})^T \boldsymbol{h_\theta}( \mathbf{y} ) | \mathbf{x} \right\} \nonumber.
\end{eqnarray}
Since $\mathbf{y}$ and ($\mathbf{z} - \mathbf{x}$) are uncorrelated or independent and two noise vectors $\mathbf{y} - \mathbf{x}$, $\mathbf{z} - \mathbf{x}$ are both zero mean, The original MSE is equivalent to the following equation:
\begin{equation}
\mathbb{E}_{(\mathbf{x}, \mathbf{y}, \mathbf{z})} \left \| \mathbf{z} - \boldsymbol{h_\theta}( \mathbf{y} ) \|^2 \right..
\label{eq:n2n}
\end{equation}
\end{proof}
Consequently, the optimal network parameters $\boldsymbol{\theta}$ of a denoiser using (\ref{eq:n2n}) will yield the same solution as the MSE based training with clean ground truth.

\subsection{Coil2Coil: generation of input and label images}

In C2C, paired noise-corrupted images are generated from the multi-coil images of phased-array coils (Fig. 1). Then, the pairs are applied to train a denoising network using N2N loss. 

First, the multi-coil images are grouped into two disjoint sets such that each set contributes to either an input image ($\hat{\mathbf{x}}_{\text{in}} \in \mathbb{R}^m$) or a label image ($\hat{\mathbf{x}}_{\text{lb}} \in \mathbb{R}^m$). Let the full set of coil indices be defined as $\mathcal{I} = \{1, \ldots, n\}$, which is divided into two subsets $\mathcal{J}, \mathcal{K} \subset \mathcal{I}$ such that $\mathcal{J} \cup \mathcal{K} = \mathcal{I}$, $\mathcal{J} \cap \mathcal{K} = \emptyset$, and $|\mathcal{J}| = |\mathcal{K}| = \frac{n}{2}$. Then the input and label images are computed as follows:

\begin{eqnarray}
\begin{split}
\hat{\mathbf{x}}_{\text{in}} &= \left| \sum_{j \in \mathcal{J}} \mathbf{s}^{(j)H} \cdot \mathbf{y}^{(j)} \right| \\
&= \left| \sum_{j \in \mathcal{J}} \mathbf{s}^{(j)H} \cdot \mathbf{s}^{(j)} \cdot \mathbf{x} + \sum_{j \in \mathcal{J}} \mathbf{s}^{(j)H} \cdot \mathbf{n}_{c}^{(j)} \right|
\end{split}
\end{eqnarray}

\begin{eqnarray}
\begin{split}
\hat{\mathbf{x}}_{\text{lb}} &= \left| \sum_{k \in \mathcal{K}} \mathbf{s}^{(k)H} \cdot \mathbf{y}^{(k)} \right| \\
&= \left| \sum_{k \in \mathcal{K}} \mathbf{s}^{(k)H} \cdot \mathbf{s}^{(k)} \cdot \mathbf{x} + \sum_{k \in \mathcal{K}} \mathbf{s}^{(k)H} \cdot \mathbf{n}_{c}^{(k)} \right|.
\end{split}
\end{eqnarray}
We assume that the two images cover the entire imaging volumes, as most of the individual coils have relatively large spatial volume coverage and mutually coupled (see Fig. 1a).

Since the combined images have reasonably high SNR such that the noise characteristics within the image can be considered as Gaussian with zero mean~\cite{ref41}, and the signal terms (e.g., $\sum_k \mathbf{s}^{(j)H} \cdot \mathbf{s}^{(j)} \cdot \mathbf{x}$ and $\sum_k \mathbf{s}^{(k)H} \cdot \mathbf{s}^{(k)} \cdot \mathbf{x}$) have real value, $\hat{\mathbf{x}}_{\text{in}}$ and $\hat{\mathbf{x}}_{\text{lb}}$ can be approximated as follows:

\begin{eqnarray}
\hat{\mathbf{x}}_{\text{in}} \simeq \sum_j |\mathbf{s}^{(j)}|^2 \cdot \mathbf{x} + \sum_{j} |\mathbf{s}^{(j)}| \cdot \mathbf{n}^{(j)}\\
\hat{\mathbf{x}}_{\text{lb}} \simeq \sum_k |\mathbf{s}^{(k)}|^2 \cdot \mathbf{x} + \sum_{k} |\mathbf{s}^{(k)}| \cdot \mathbf{n}^{(k)}
\end{eqnarray}
where \(\mathbf{n}^{(i)}  \in \mathbb{R}^m\) is the noise, modeled as zero-mean Gaussian noise with a standard deviation of \(\sigma^{(i)}\). These two images, however, have different coil sensitivity weighting (see $\hat{\mathbf{x}}_{\text{in}}$ and $\hat{\mathbf{x}}_{\text{lb}}$ in Fig. 1b), resulting in a different clean image, and may have correlated noise (e.g., mutual inductance between channels)~\cite{ref42}. Therefore, they need to be further processed to using N2N loss. First, to decorrelate the noise,  $\hat{\mathbf{x}}_{\text{lb}}$ was revised:

\begin{lemma}
Given $\hat{\mathbf{x}}_{\text{in}}$ and $\hat{\mathbf{x}}_{\text{lb}}$ which is correlated, we can construct $\hat{\mathbf{x}}^{'}_{\text{lb}}$ where $\hat{\mathbf{x}}_{\text{in}}$ and $(\hat{\mathbf{x}}^{'}_{\text{lb}} - \sum_j | \mathbf{s}^{(j)}|^2 \cdot \mathbf{x})$ are uncorrelated using a generalized least-square solution~\cite{ref43}:

\begin{equation}
    \hat{\mathbf{x}}^{'}_{\text{lb}} = \boldsymbol{\alpha} \cdot \hat{\mathbf{x}}_{\text{in}} + \boldsymbol{\beta} \cdot \hat{\mathbf{x}}_{\text{lb}},
\end{equation}
where the coefficients $\boldsymbol{\alpha}$ and $\boldsymbol{\beta}$ are defined as:

\begin{equation}
\boldsymbol{\alpha} = \frac{-\boldsymbol{\sigma}_{JK}^2}
{\sqrt{\boldsymbol{\sigma}_J^2 \cdot \boldsymbol{\sigma}_K^2 - \left(\boldsymbol{\sigma}_{JK}^2\right)^2}},
\end{equation}

\begin{equation}
\boldsymbol{\beta} = \frac{\boldsymbol{\sigma}_{J}^2}
{\sqrt{\boldsymbol{\sigma}_J^2 \cdot \boldsymbol{\sigma}_K^2 - \left(\boldsymbol{\sigma}_{JK}^2\right)^2}},
\end{equation}
and $\boldsymbol{\sigma}_{J}^2 = \mathrm{cov}(\hat{\mathbf{x}}_{\text{in}})$, $\boldsymbol{\sigma}_{K}^2 = \mathrm{cov}(\hat{\mathbf{x}}_{\text{lb}})$, and $\boldsymbol{\sigma}_{JK}^2 = \mathrm{cov}(\hat{\mathbf{x}}_{\text{in}}, \hat{\mathbf{x}}_{\text{lb}})$ are vectors in $\mathbb{R}^m$.
\end{lemma}

\begin{proof}
The goal is to construct $\hat{\mathbf{x}}^{'}_{\text{lb}}$ such that:
\[
\mathrm{cov}\left( \hat{\mathbf{x}}_{\text{in}}, \hat{\mathbf{x}}^{'}_{\text{lb}} - \sum_j |\mathbf{s}^{(j)}|^2 \cdot \mathbf{x} \right) = 0.
\]
Substituting the expression for $\hat{\mathbf{x}}^{'}_{\text{lb}}$ gives:
\begin{align*}
&\text{cov} \left(\hat{\mathbf{x}}_{\text{in}}, \boldsymbol{\alpha} \cdot \hat{\mathbf{x}}_{\text{in}} + \boldsymbol{\beta} \cdot \hat{\mathbf{x}}_{\text{lb}} - \sum_j |\mathbf{s}^{(j)}|^2 \cdot \mathbf{x} \right) \\
&= \boldsymbol{\alpha} \cdot \text{cov}(\hat{\mathbf{x}}_{\text{in}}) + \boldsymbol{\beta} \cdot \text{cov}(\hat{\mathbf{x}}_{\text{in}}, \hat{\mathbf{x}}_{\text{lb}}) \\
&= \boldsymbol{\alpha} \cdot \boldsymbol{\sigma}_J^2 + \boldsymbol{\beta} \cdot \boldsymbol{\sigma}_{JK}^2.
\end{align*}
Applying the definitions of $\boldsymbol{\alpha}$ and $\boldsymbol{\beta}$:
\[
\boldsymbol{\alpha} \cdot \boldsymbol{\sigma}_J^2 + \boldsymbol{\beta} \cdot \boldsymbol{\sigma}_{JK}^2 = 
\frac{-\boldsymbol{\sigma}_{JK}^2 \cdot \boldsymbol{\sigma}_J^2 + \boldsymbol{\sigma}_{J}^2 \cdot \boldsymbol{\sigma}_{JK}^2}
{\sqrt{\boldsymbol{\sigma}_J^2 \cdot \boldsymbol{\sigma}_K^2 - \left( \boldsymbol{\sigma}_{JK}^2 \right)^2}} = 0.
\]
Thus, the constructed $\hat{\mathbf{x}}^{'}_{\text{lb}}$ satisfies the uncorrelated condition.
\end{proof}

After the decorrelation, $\hat{\mathbf{x}}^{'}_{\text{lb}}$ was further processed to match the underlying clean image with $\hat{\mathbf{x}}_{\text{in}}$:

\begin{lemma}
Given $\hat{\mathbf{x}}_{\text{in}}$ and $\hat{\mathbf{x}}^{'}_{\text{lb}}$, $\hat{\mathbf{x}}''_{\text{lb}}$ can be constructed such that it shares the same clean component (=$\sum_j |\mathbf{s}^{(j)}|^2 \cdot \mathbf{x}$) as $\hat{\mathbf{x}}_{\text{in}}$:
\begin{equation}
\hat{\mathbf{x}}''_{\text{lb}} = \left( \mathbf{s}_{\text{input}} \oslash \mathbf{s}'_{\text{label}} \right) \cdot \hat{\mathbf{x}}'_{\text{lb}},
\end{equation}
where $\oslash$ denotes the Hadamard (element-wise) division, where the sensitivity maps $\mathbf{s}'_{\text{label}}$ and $\mathbf{s}_{\text{input}}$ are defined as:
\begin{align}
\mathbf{s}'_{\text{label}} &= \boldsymbol{\alpha} \cdot \sum_j |\mathbf{s}^{(j)}|^2 + \boldsymbol{\beta} \cdot \sum_k |\mathbf{s}^{(k)}|^2, \\
\mathbf{s}_{\text{input}} &= \sum_j |\mathbf{s}^{(j)}|^2.
\end{align}
\end{lemma}
\begin{proof}
A scaling ratio $\boldsymbol{\gamma}$ can be defined as:
\begin{equation}
\boldsymbol{\gamma} = \mathbf{s}_{\text{input}} \oslash \mathbf{s}_{\text{label}}^{'}.
\end{equation}
Applying the scaling to $\hat{\mathbf{x}}^{'}_{\text{lb}}$ yields:
\begin{equation}
\begin{split}
\hat{\mathbf{x}}^{''}_{\text{lb}}& = \sum_j |\mathbf{s}^{(j)}|^2 \cdot \mathbf{x} \\
&+ \boldsymbol{\gamma} \cdot (\boldsymbol{\alpha} \cdot \sum_j |\mathbf{s}^{(j)}| \cdot \mathbf{n}^{(j)}+ \boldsymbol{\beta} \cdot \sum_k |\mathbf{s}^{(k)}| \cdot \mathbf{n}^{(k)}).
\end{split}
\end{equation}
Thus, $\hat{\mathbf{x}}''_{\text{lb}}$ shares the same clean signal $\sum_j |\mathbf{s}^{(j)}|^2 \cdot \mathbf{x}$ as $\hat{\mathbf{x}}_{\text{in}}$.
\end{proof}

Since the transformation from $\hat{\mathbf{x}}^{'}_{\text{lb}}$ to $\hat{\mathbf{x}}^{''}_{\text{lb}}$ is performed via an element-wise operation that only scales each component, the uncorrelated property between $\hat{\mathbf{x}}_{\text{in}}$ and $\hat{\mathbf{x}}^{''}_{\text{lb}} - \sum_j |\mathbf{s}^{(j)}|^2 \cdot \mathbf{x}$, established in \textbf{Lemma 2}, is maintained. Consequently, training a network with the pair $(\hat{\mathbf{x}}_{\text{in}}, \hat{\mathbf{x}}^{''}_{\text{lb}})$ is equivalent to  training with the clean target, as stated in the following theorem:

\begin{theorem}
Training a denoising network $\boldsymbol{h_\theta}$ using the pair $(\hat{\mathbf{x}}_{\text{in}}, \hat{\mathbf{x}}^{''}_{\text{lb}})$ is equivalent (in expectation) to training with the clean target:
\begin{equation}
\begin{split}
\mathbb{E}&_{(\sum_j |\mathbf{s}^{(j)}|^2 \cdot \mathbf{x},\hat{\mathbf{x}}^{''}_{\text{lb}}, \hat{\mathbf{x}}_{\text{in}})} \| \hat{\mathbf{x}}^{''}_{\text{lb}} - \boldsymbol{h_\theta}(\hat{\mathbf{x}}_{\text{in}}) \|^2 \\
&=\mathbb{E}_{(\sum_j |\mathbf{s}^{(j)}|^2 \cdot \mathbf{x}, \hat{\mathbf{x}}_{\text{in}})} \| \sum_j |\mathbf{s}^{(j)}|^2 \cdot \mathbf{x} - \boldsymbol{h_\theta}(\hat{\mathbf{x}}_{\text{in}}) \|^2.
\end{split}
\end{equation}
\end{theorem}

\begin{proof}
Consider the triplet $(\sum_j |\mathbf{s}^{(j)}|^2 \cdot \mathbf{x}, \hat{\mathbf{x}}_{\text{in}}, \hat{\mathbf{x}}^{''}_{\text{lb}})$ in the context of the general triplet $(\mathbf{x}, \mathbf{y}, \mathbf{z})$ in \textbf{Lemma 1}. Then, we observe that $\mathbf{y} - \mathbf{x}$, $\mathbf{z} - \mathbf{x}$ are zero mean vectors, establised in \textbf{Lemma 3}, and $\mathbf{y}$ and ($\mathbf{z} - \mathbf{x}$) are uncorrelated, as shown in \textbf{Lemma 2}. Therefore, Training with $(\mathbf{i}_{\text{input}}, \mathbf{i}_{\text{label}}^{''})$ yields the same expected loss as training with the clean signal $\sum_j |\mathbf{s}^{(j)}|^2 \cdot \mathbf{x}$.
\end{proof}

This loss function is calculated within a brain mask. For inference, the combined image, $\hat{\mathbf{x}}_{\text{cb}}$, which combined all channel images, was used as an input image (Fig. 1c). Finally, scanner-generated DICOM images, which can also be regarded as combined image orignated from scanner, are also evaluated.

\begin{table}[!t]
\caption{pSNR and SSIM of the denoising methods for the synthetic noise denoising experiment. A statistically significant difference exists between all self-supervised learning methods and C2C.}
\centering
\resizebox{\linewidth}{!}{ 
\begin{tabular}{c c c c}
\hline\hline
Type & Method & pSNR & SSIM \\
\hline
Supervised learning & N2C & 39.62 ± 2.69	& 0.979 ± 0.010 \\
\hline
\multirow{6}{*}{\begin{tabular}{c}Self-supervised\\learning\end{tabular}} 
& C2C (Ours)  & 39.01 ± 2.91  & 0.969 ± 0.017 \\
& N2V  & 36.32 ± 3.20 & 0.934 ± 0.031 \\
& N2Se & 37.40 ± 3.16 & 0.952 ± 0.027 \\
& N2Sa & 37.54 ± 2.74 & 0.954 ± 0.021 \\
& Ne2Ne  & 37.81 ± 2.44  & 0.958 ± 0.022 \\
& N2N*  & 36.53 ± 2.48  & 0.948 ± 0.041 \\
\hline\hline
\end{tabular}
}
\end{table}

\subsection{Deep neural networks}
For the structure of the neural network, NAFNet~\cite{ref44}, was applied. The initial weights of the network parameters were set by the Xavier initializer~\cite{ref45}. The learning rate was 1e-4, and a decaying factor of 0.87 was applied for each epoch. For an optimizer, an Adam optimizer was utilized~\cite{ref46}. The batch size was 8. The training process was stopped after 100 epochs. For each epoch, the channel combinations for the input and label images were randomly determined for generalization.

The network training was performed on a GPU workstation (TITAN Xp GPU with Intel i7-7800X CPU at 3.50GHz) using PyTorch~\cite{ref47}. The total training time was approximately 86 hours.

\subsection{Datasets}
For the training and test, the NYU fastMRI brain train and test datasets, containing T1-weighted, T2-weighted, and fluid attenuation inversion recovery (FLAIR) images from 1.5 T and 3 T were utilized~\cite{ref48}. The number of channels ranged from 4 to 20. 

As a preprocessing step, each subject was normalized by the mean and standard deviation. The upper three slices were excluded because they were usually background. A total of 34341 and 8284 slices were utilized as the training and test datasets, respectively. The sensitivity map of each channel image was calculated using ESPIRiT~\cite{ref40}. The brain mask was obtained using a brain extraction tool~\cite{ref49}. All preprocessing steps were performed using MATLAB (MATLAB2020a, MathWorks Inc., Natick, MA, USA).

\begin{figure*}[!t]
\centering
\includegraphics[width=1\textwidth]{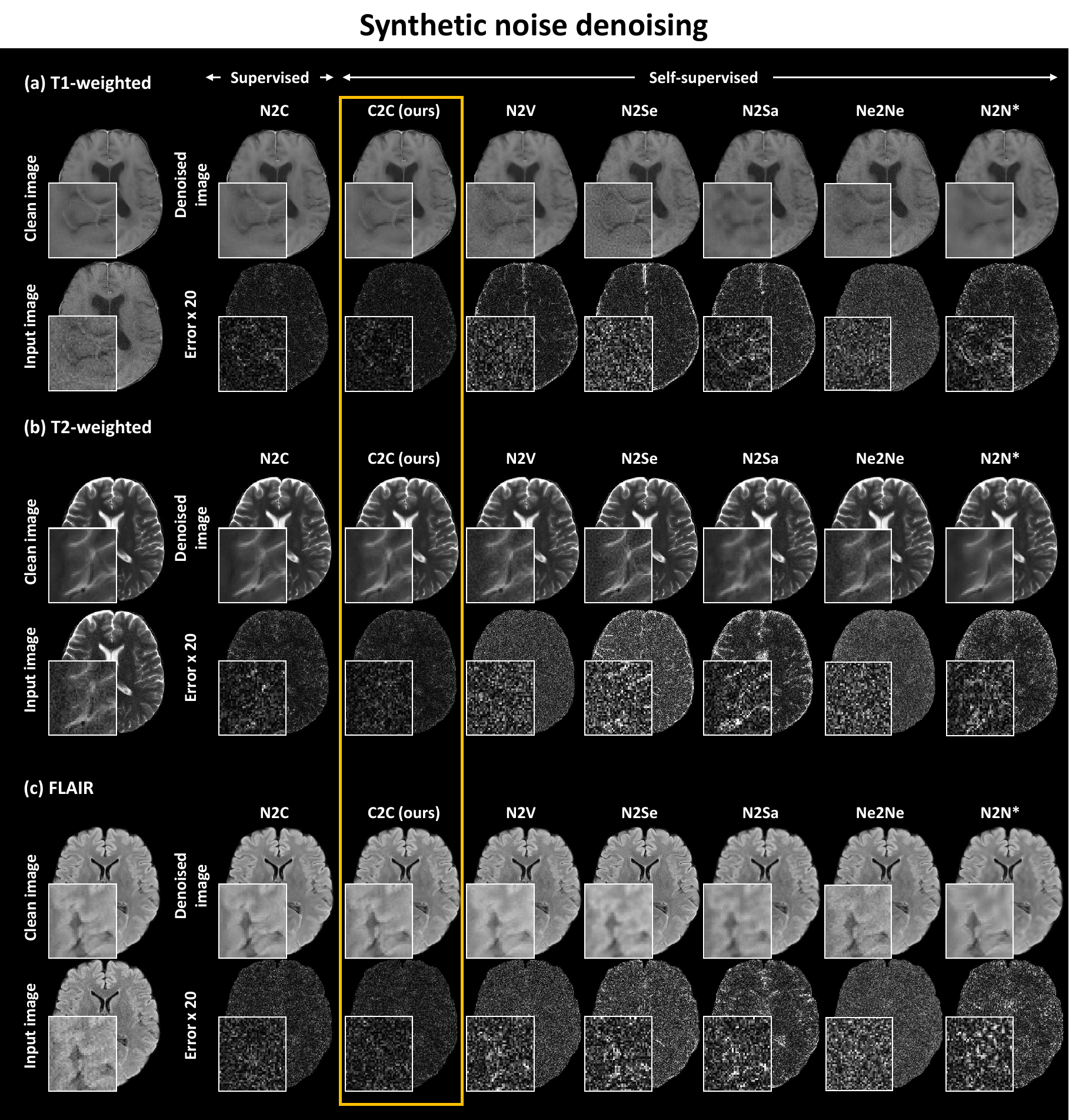}
\caption{Denoised images of the synthetic noise denoising experiment. (a) T1-weighted images: The clean image and noise-corrupted input image are shown in the first column. Denoised images using N2C, C2C, N2V, N2Se, N2Sa, Ne2Ne, and N2N* are shown along with the error maps at the scale of error × 20. The results of T2-weighted images (b) and FLAIR images (c) are presented in the same format. C2C provides comparable outcomes to N2C that require additional acquisition of paired images for training, while showing superior images to those of the other self-supervised learning methods.}
\label{fig_2}
\end{figure*}

\subsection{Denoising of synthetic noise added images}
To quantitatively analyze the performance of C2C, denoising experiments were performed by simulating the noise-corrupted image and corresponding clean image. The combined image $\hat{\mathbf{x}}_{\text{cb}}$ was assumed to be a clean image in this synthetic noise experiment. The multi-coil images were simulated by multiplying the coil sensitivity $\mathbf{s}^{(i)}$ to $\hat{\mathbf{x}}_{\text{cb}}$, and then corrupted by complex noise ($\mathbf{n}_{r}^{(i)} + i \mathbf{n}_{i}^{(i)}$; each axis with zero-mean Gaussian noise with the standard deviation of $\overline{|\mathbf{s}^{(i)} \cdot \hat{\mathbf{x}}_{\text{cb}}|} / \sqrt{2}$
where $\overline{|\mathbf{s}^{(i)} \cdot \hat{\mathbf{x}}_{\text{cb}}|}$ was the mean intensity within the brain mask).

The performance of C2C was compared to that of other denoising methods. As a supervised method, Noise2Clean (N2C)~\cite{ref22}, which used a noise-corrupted image as an input and a clean image as a label, was tested. For self-supervised methods, N2V, N2Se, N2Sa, Ne2Ne, and N2N* were tested. These self-supervised methods utilized a single noise-corrupted image $\hat{\mathbf{x}}_{\text{cb}}$ for training, with the same network structure as in C2C. In order to evaluate the denoising performance, the quantitative metrics, peak signal-to-noise ratio (pSNR) and structural similarity index (SSIM), were measured between the denoised image and clean image within the brain mask. The means and standard deviation of the quantitative metrics were reported for all slices of the test dataset. A paired $t$-test was performed for the quantitative metrics between C2C and the self-supervised methods. For statistical significance, the $p$-value threshold was set to 0.05.

\begin{table}[!t]
\caption{Quantitative metrics in the inter-channel correlated noise denoising experiment. In NaiveC2C, noise uncorrelation process was not implemented.}
\centering
\resizebox{\linewidth}{!}{ 
\begin{tabular}{c c c c}
\hline\hline
Type & Method & pSNR & SSIM \\
\hline
Supervised learning & N2C & 39.55 ± 2.68	& 0.978 ± 0.011 \\
\hline
\multirow{7}{*}{\begin{tabular}{c}Self-supervised\\learning\end{tabular}} 
& C2C (Ours)  & 38.94 ± 2.92  & 0.970 ± 0.015 \\
& NaiveC2C  & 36.88 ± 2.90  & 0.941 ± 0.030 \\
& N2V  & 36.22 ± 3.21 & 0.932 ± 0.034 \\
& N2Se & 37.35 ± 3.17 & 0.952 ± 0.022 \\
& N2Sa & 37.40 ± 2.74 & 0.955 ± 0.024 \\
& Ne2Ne  & 37.68 ± 2.42  & 0.960 ± 0.021 \\
& N2N*  & 36.50 ± 2.50  & 0.944 ± 0.035 \\
\hline\hline
\end{tabular}
}
\end{table}

\subsection{Denoising of inter-channel or spatially correlated noise}
To explore the denoising performance when various types of correlated noise exists, ablation studies were performed. Two types of correlated noise were tested: inter-channel correlation and spatial correlation. 

First, the effects of inter-channel correlation were explored. Instead of adding independent Gaussian noise, the noise across channels was modeled as a multivariate Gaussian distribution. For each channel \( i \), the noise \(\mathbf{n}_{c}^{(i)} = \mathbf{n}_{r}^{(i)} + i \mathbf{n}_{i}^{(i)}\) was generated from:
\[
(n_c^{(1)}, n_c^{(2)}, \dots, n_c^{(n)}) \sim \mathcal{CN}(0, \mathbf{\Sigma})
\]
where \( \mathbf{\Sigma} \) is the covariance matrix.
The diagonal elements were set as equal as in the independent experiment, while the off-diagonal elements were set to 0.3, which is sufficiently high to cover the range of inter-coil correlations in practice~\cite{ref50}.
N2C and self-supervised methods were re-trained and tested. Additionally, to explore the effects of the noise uncorrelation process in C2C, NaiveC2C was tested, where the noise uncorrelation process was not implemented.

Second, the effects of spatial correlation in noise were explored. A spatially correlated noise was generated by a kernel from~\cite{ref35} with \(\beta = 0.5\). N2C and self-supervised methods were re-trained and tested.

\begin{table}[!t]
\caption{Quantitative metrics in the spatially correlated noise denoising experiment.}
\centering
\resizebox{\linewidth}{!}{ 
\begin{tabular}{c c c c}
\hline\hline
Type & Method & pSNR & SSIM \\
\hline
Supervised learning & N2C & 39.57 ± 2.69	& 0.974 ± 0.011 \\
\hline
\multirow{6}{*}{\begin{tabular}{c}Self-supervised\\learning\end{tabular}} 
& C2C (Ours)  & 38.61 ± 2.91  & 0.964 ± 0.020 \\
& N2V  & 33.75 ± 3.20 & 0.907 ± 0.051 \\
& N2Se & 33.88 ± 3.16 & 0.910 ± 0.040 \\
& N2Sa & 34.40 ± 2.74 & 0.911 ± 0.047 \\
& Ne2Ne  & 34.62 ± 2.44  & 0.919 ± 0.039 \\
& N2N*  & 36.17 ± 2.48  & 0.941 ± 0.042 \\
\hline\hline
\end{tabular}
}
\end{table}

\subsection{Denoising of real world images}

Experiments were performed to denoise real-world images (i.e., DICOM images given directly  from scanners). The C2C network was trained using the real noise-corrupted multi-coil images. For the test, the NYU fastMRI brain DICOM dataset, containing T1-weighted, T2-weighted, and FLAIR images, was utilized~\cite{ref49}. For comparison, the self-supervised methods, N2V, N2Se, N2Sa, Ne2Ne, and N2N* were trained and tested using the same datasets.

\begin{figure*}[!t]
\centering
\includegraphics[width=1\textwidth]{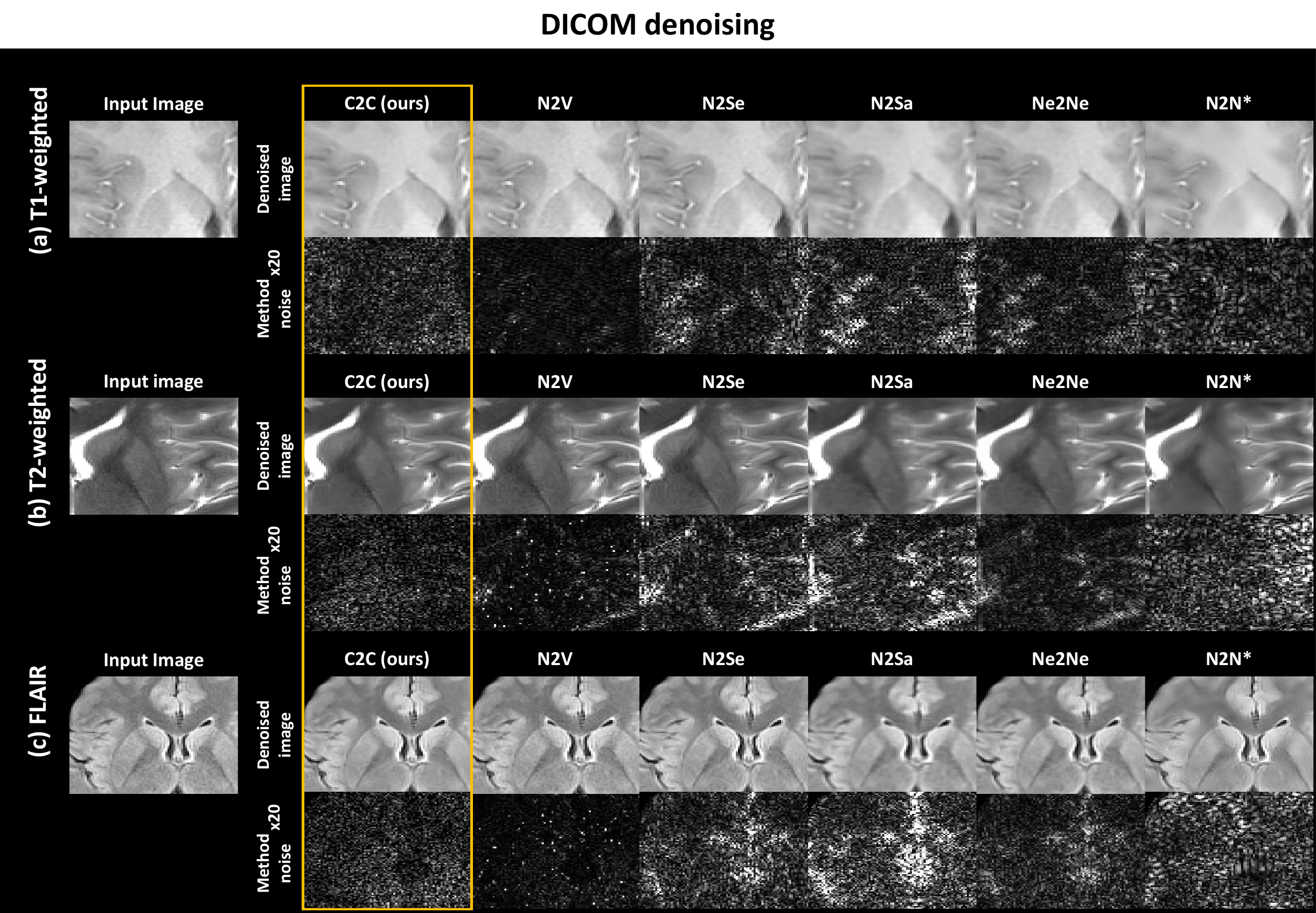}
\caption{Denoising results of the real-world DICOM images using the self-supervised methods. (a) T1-weighted image: The input image contains a noticeable level of noise whereas the denoised images using C2C, N2V, N2Se, N2Sa, Ne2Ne, and N2N* show improved results. Method noise (i.e., the difference between the input and the denoised image) is displayed below the corresponding denoised images. The results of T2-weighted images (b) and FLAIR images (c) are also displayed. In all results, C2C images are superior to the other results, showing little or no structure in the method noise.}
\label{fig_5}
\end{figure*}

\section{Results}

The results of the synthetic noise denoising experiment are shown in Fig. 2. When we denoised the synthetic noise added images, our method, C2C, generated high-quality images similar to the clean images in all three contrasts (see the images in the yellow boxes and compared them to the clean images in the first column). When compared to the results of N2C, the C2C images reveal similar image qualities that can be confirmed by the error × 20 images. On the other hand, the outputs of the other self-supervised learning methods show performance degradations that are manifested in both output and error images (e.g., persistent noise in the N2V, N2Se, and Ne2Ne images, whereas blurring in the N2Sa and N2N* images). When analyzed quantitatively, the performance of C2C is comparable to those of N2C despite the fact C2C does not require acquisition of clean or paired images for training (Table 1). Furthermore, C2C reports the highest pSNR (39.01 ± 2.91) and SSIM (0.968 ± 0.017) among the rest of the self-supervised learning methods, confirming the visual inspection results (Table 1). 

When evaluating the effects of the inter-channel noise correlation, C2C still shows the performance comparable to N2C (Table 2), while outperforming the other self-supervised denoising methods. Furthermore, the performance of NaiveC2C demonstrates that the noise uncorrelation process improves the denoising quality. In the experiment of the spatially correlated noise denoising, C2C successfully shows the high denoising performance, comparable to the N2C, while surpassing the other self-supervised denoising methods.

When the real-world images (i.e., DICOM) are processed for denoising (Fig. 3), C2C substantially improved image quality, showing little or no structure-dependent error (method noise~\cite{ref4}; yellow box). Similar to the synthetic noise experiment, N2V, N2Se, and Ne2Ne images yield less denoised images while N2Sa and N2N* images show blurring. No results of N2C are available because the dataset contains no clean image nor paired noise-corrupted image.

\section{Discussion}

In this study, we proposed a novel self-supervised MRI denoising framework, C2C, which generates the pair of noise-corrupted images from multi-coil images to train a denoising network. By exploiting the multi-coil images, our method produces realistic noise-corrupted image pairs, achieving performance comparable to supervised learning while outperforming other self-supervised methods. Furthermore, C2C demonstrated robustness against the inter-channel or spatially correlated noise. Our method successfully denoised the real-world images, showing no structure-dependency in the denoised map, demonstrating wide applicability.  

In previous studies, a few self-supervised denoising methods were proposed to utilize redundancy in data from specific sequences (e.g., DWI~\cite{ref36, ref37} and dynamic MR~\cite{ref38}), limiting applications of the methods. On the other hand, C2C is not limited to a specific sequence and, therefore, can be utilized more generally. In addition, since the modern MRI system mostly acquiring MR image using phased-array coil system, our method was not limited by hardware. Furthermore, it can be extended to train complex images in order to denoise both magnitude and phase images. 
	
In C2C training, the noise level of the paired images can be different depending on the combinations of channel images or the scaling process. This does not violate the conditions of N2N because the conditions do not require the same level of noise in the paired images~\cite{ref23}. 
	
In MRI, noise originates from multiple sources such as samples (e.g., brain) and hardware (e.g., receiver chain)~\cite{ref51}. Since sample noise is common in all channels, we expect our method removes noise from the receiver chain (e.g., coil, preamplifier, etc.), which may or may not be dominant depending on the size of the coil relative to the sample~\cite{ref52}. Furthermore, the noise model used in C2C is based on several assumptions. The k-space is assumed to be sampled in a Cartesian coordinate with FFT reconstruction, excluding reconstruction methods such as parallel imaging and compressed sensing. Furthermore, post-processing such as apodization, interpolation and pre-scan normalization would also affect the noise modeling. Despite the limitations, the real-world data results indicate that the method may perform beyond the receiver chain noise removal, even in DICOM data that deviates from the noise model, potentially benefiting from the nature of a denoising neural network that inputs multiple voxels, utilizing information from neighboring voxels for denoising.

As an alternative option for inference, we tested an approach that combined all channel images into two and then inferred the two images, generating two C2C denoised images. When these images were averaged as the final output, the performances were slightly worse than the current approach of combining all channels and then inferring (pSNR dropped from 38.97 ± 2.91 to 38.10 ± 2.88 and SSIM reduced from 0.968 ± 0.017 to 0.962 ± 0.017). This result might be explained by a lower SNR of the partially coil-combined images than that of a fully coil-combined image.
	
 The proposed method is based on the assumption that the paired images generated from different combinations of the channels sufficiently cover the imaging volume which may not be valid when using a coil with a small number of channels or channels with limited coverages, potentially degrading the performance of our method. For example, when coil elements were reduced from 20 to 10 or 5 for the 20 channel coil test dataset, pSNR changed from 39.71 ± 1.81 to 36.57 ± 2.25 for 10 channels or 31.41 ± 2.84 for 5 channels, demonstrating the substantial effects of the coil coverage. 
	
 Our method utilized the redundancy of phased-array coil data. It may be combined with additional redundancy (e.g., multi-echo data, multi-contrast data) to provide an further improvement in denoising performance.
 
\section{Conclusion}

In this paper, we proposed C2C, a self-supervised denoising framework for MRI that leverages the phased-array coil data to generate noise-corrupted image pairs without clean images. The method ensures statistical independence of the two noise-corrupted images while sharing the same underlying clean image. 

\section*{Acknowledgments}
This research was supported by the National Research Foundation of Korea grant funded by MSIT (No. NRF-2022R1A4A1030579), Ministry of Education (RS-2024-00435727), Korea health technology R\&D project through the KHIDI, funded by the Ministry of Health \& Welfare (RS-2024-00439677), Samsung Electronics Co., Ltd (IO201216-08215-01), Institute of Engineering Research, and Institute of New Media and Communications at Seoul National University.

\vfill


\begin{thebibliography}{1}
\bibliographystyle{IEEEtran}
\bibitem{ref1}
A. Macovski, ``Noise in MRI,'' \textit{Magn. Reson. Med.}, vol. 36, no. 3, pp. 494--497, Sep. 1996.
\bibitem{ref2}
J. Mohan, V. Krishnaveni, and Y. Guo, ``A survey on the magnetic resonance image denoising methods,'' \textit{Biomed. Signal Process. Control}, vol. 9, pp. 56--69, Jan. 2014.
\bibitem{ref3}
K. Dabov, A. Foi, V. Katkovnik, and K. Egiazarian, ``Image denoising by sparse 3-D transform-domain collaborative filtering,'' \textit{IEEE Trans. Imag. Proc.}, vol. 16, no. 8, pp. 2080--2095, Jul. 2007.
\bibitem{ref4}
A. Buades, B. Coll, and J.-M. Morel, ``A non-local algorithm for image denoising,'' in \textit{Proc. IEEE Int. Conf. Comput. Vis. Pattern Recognit.}, vol. 2, pp. 60--65, Jul. 2005.
\bibitem{ref5}
T. Loupas, W. McDicken, and P. L. Allan, ``An adaptive weighted median filter for speckle suppression in medical ultrasonic images,'' \textit{IEEE Trans. Circuit. Sys.}, vol. 36, no. 1, pp. 129--135, Jan. 1989.
\bibitem{ref6}
S. V. M. Sagheer and S. N. George, ``Ultrasound image despeckling using low rank matrix approximation approach,'' \textit{Biomed. Signal Process. Control}, vol. 38, pp. 236--249, Sep. 2017.
\bibitem{ref7}
H. S. Khaleel, S. V. M. Sagheer, M. Baburaj, and S. N. George, ``Denoising of Rician corrupted 3D magnetic resonance images using tensor-SVD,'' \textit{Biomed. Signal Process. Control}, vol. 44, pp. 82--95, Jul. 2018.
\bibitem{ref8}
C. Tian, L. Fei, W. Zheng, Y. Xu, W. Zuo, and C.-W. Lin, ``Deep learning on image denoising: An overview,'' \textit{Neural Networks}, vol. 131, pp. 251--275, Nov. 2020.
\bibitem{ref9}
H. C. Burger, C. J. Schuler, and S. Harmeling, ``Image denoising: Can plain neural networks compete with BM3D?,'' in \textit{Proc. IEEE Int. Conf. Comput. Vis. Pattern Recognit}, vol. 2, pp. 2392--2399, Jun. 2012.
\bibitem{ref10}
A. S. Lundervold and A. Lundervold, ``An overview of deep learning in medical imaging focusing on MRI,'' \textit{Zeitschrift Für Medizinische Physik}, vol. 29, pp. 102--127, May 2019.
\bibitem{ref11}
M. Akçakaya, S. Moeller, S. Weingärtner, and K. Uğurbil, ``Scan‐specific robust artificial‐neural‐networks for k‐space interpolation (RAKI) reconstruction: Database‐free deep learning for fast imaging,'' \textit{Magn. Reson. Med.}, vol. 81, pp. 439--453, Sep. 2019.
\bibitem{ref12}
J. Yoon et al., ``Quantitative susceptibility mapping using deep neural network: QSMnet,'' \textit{Neuroimage}, vol. 179, pp. 199--206, Oct. 2018.
\bibitem{ref13}
D. Shin et al., ``Deep reinforcement learning-designed radiofrequency waveform in MRI,'' \textit{Nature Mach. Intel.}, vol. 3, no. 11, pp. 985--994, Nov. 2021.
\bibitem{ref14}
L. Gondara, ``Medical image denoising using convolutional denoising autoencoders,'' in \textit{Proc Int. Conf. data mining workshops}, pp. 241--246, Dec. 2016.
\bibitem{ref15}
D. Jiang, W. Dou, L. Vosters, X. Xu, Y. Sun, and T. Tan, ``Denoising of 3D magnetic resonance images with multi-channel residual learning of convolutional neural network,'' \textit{Jpn. J. Radiol.}, vol. 36, no. 9, pp. 566--574, Jul. 2018.
\bibitem{ref16}
X. Xu et al., ``Noise Estimation-based Method for MRI Denoising with Discriminative Perceptual Architecture,'' in \textit{Proc. Int. Conf. Internet of Things (iThings) IEEE Green Comput. and Commun. (GreenCom) and IEEE Cyber Phys. Soc. Comput. (CPSCom) and IEEE Smart Data (SmartData) IEEE Congr. Cybermatics (Cybermatics)}, pp. 469--473, Nov. 2020.
\bibitem{ref17}
J. V. Manjón and P. Coupe, ``MRI denoising using deep learning,'' in \textit{Proc. Int. Patch Based Tech. Med. Imag.}, pp. 12--19, Sep. 2018.
\bibitem{ref18}
S. Li, J. Zhou, D. Liang, and Q. Liu, ``MRI denoising using progressively distribution-based neural network,'' \textit{Magn. Reson. Med.}, vol. 71, pp. 55--68, Sep. 2020.
\bibitem{ref19}
M. Kidoh et al., ``Deep learning based noise reduction for brain MR imaging: tests on phantoms and healthy volunteers,'' \textit{Magn. Reson. Med. Sci.}, vol. 19, no. 3, pp. 195--206, Sep. 2020.
\bibitem{ref20}
T. Higaki et al., ``Improvement of image quality at ct and mri using deep learning,'' \textit{Jpn. J. Radiol.}, vol. 37, no. 1, pp. 73--80, Jan. 2019.
\bibitem{ref21}
C. Bermudez et al., ``Learning implicit brain MRI manifolds with deep learning,'' in \textit{Proc. SPIE Medical Imaging: Image Processing}, vol. 10574, pp. 105741L, Mar. 2018.
\bibitem{ref22}
V. Jain and S. Seung, ``Natural image denoising with convolutional networks,'' in \textit{Proc. Adv. Neural Inf. Process. Syst.}, vol. 21, pp. 769--776, Dec. 2008.
\bibitem{ref23}
J. Lehtinen et al., ``Noise2Noise: Learning image restoration without clean data,'' in \textit{Proc. Int. Conf. Mach. Learn.}, pp. 2965--2974, Jul. 2018.
\bibitem{ref24}
A. A. Hendriksen, D. M. Pelt, and K. J. Batenburg, ``Noise2Inverse: Self-supervised deep convolutional denoising for tomography,'' \textit{IEEE Trans. Comput. Imag.}, vol. 6, pp. 1320--1335, Aug. 2020.
\bibitem{ref25}
A. Krull, T.-O. Buchholz, and F. Jug, ``Noise2Void-learning denoising from single noisy images,'' in \textit{Proc. IEEE Int. Conf. Comput. Vis. Pattern Recognit.}, pp. 2129--2137, Jun. 2019.
\bibitem{ref26}
J. Batson and L. Royer, ``Noise2Self: Blind denoising by self-supervision,'' in \textit{Proc. Int. Conf. Mach. Learn.}, pp. 524--533, Jun. 2019.
\bibitem{ref27}
Y. Xie, Z. Wang, and S. Ji, ``Noise2Same: Optimizing a self-supervised bound for image denoising,'' in \textit{Proc. Adv. Neural Inf. Process. Syst.}, vol. 33, pp. 20320--20330, Dec. 2020.
\bibitem{ref28}
T. Huang, S. Li, X. Jia, H. Lu, J. Liu, ``Neigbor2Neighbor: Self-supervised denoising from single noisy images,'' in \textit{Proc. IEEE Int. Conf. Comput. Vis. Pattern Recognit.}, pp. 14781--14790, Jun. 2021.
\bibitem{ref29}
M. Zhussip, S. Soltanayev, and S. Y. Chun, ``Extending Stein's unbiased risk estimator to train deep denoisers with correlated pairs of noisy images,'' in \textit{Proc. Adv. Neural Inf. Process. Syst.}, vol. 32, pp. 1465--1475, Dec. 2019.
\bibitem{ref30}
S. Soltanayev and S. Y. Chun, ``Training deep learning based denoisers without ground truth data,'' in \textit{Proc. Adv. Neural Inf. Process. Syst.}, vol. 31, pp. 3261--3271, Dec. 2018.
\bibitem{ref31}
N. Moran, D. Schmidt, Y. Zhong and P. Coady, ``Noisier2noise: Learning to denoise from unpaired noisy data,'' in \textit{Proc. IEEE Int. Conf. Comput. Vis. Pattern Recognit.}, pp. 12064--12072, Jun. 2020.
\bibitem{ref32}
K. Kim and J.C.Ye, ``Noise2score: tweedie’s approach to self-supervised image denoising without clean images,'' in \textit{Proc. Adv. Neural Inf. Process. Syst.}, vol. 34, pp. 864--874, Dec. 2021.
\bibitem{ref33}
Yuhui Quan, Mingqin Chen, Tongyao Pang, and Hui Ji. Self2self with dropout: Learning self-supervised denoising from single image. In \textit{CVPR}, pages 1890–1898, 2020.
\bibitem{ref34}
Zejin Wang, Jiazheng Liu, Guoqing Li, and Hua Han. Blind2unblind: Self-supervised image denoising with visible blind spots. \textit{In CVPR}, pages 2027-2036, 2022.
\bibitem{ref35}
H. Chihaoui and P. Favaro, “Masked and shuffled blind spot denoising for real-world images,” \textit{in Proceedings of the IEEE/CVF Conference on Computer Vision and Pattern Recognition}, 2024, pp. 3025–3034.
\bibitem{ref36}
S. Fadnavis, J. Batson, and E. Garyfallidis, ``Patch2Self: denoising diffusion MRI with self-supervised learning,'' in \textit{Proc. Adv. Neural Inf. Process. Syst.}, vol. 33, pp. 1--11, Dec. 2020.
\bibitem{ref37}
Q. Tian, Z. Li, Q. Fan, J. R. Polimeni, B. Bilgic, D. H. Salat, and
S. Y. Huang, “Sdndti: Self-supervised deep learning-based denoising
for diffusion tensor mri,” \textit{Neuroimage}, vol. 253, p. 119033, 2022
\bibitem{ref38}
J. Xu and E. Adalsteinsson, ``Deformed2Self: Self-Supervised Denoising for Dynamic Medical Imaging,'' in \textit{Proc. Int. Conf. Med. Image Comput. Comput.Assist. Intervent.}, vol. 10902, pp. 25--35, Sep. 2021.
\bibitem{ref39}
Lustig, M., Donoho, D. L., Santos, J. M., Pauly, J. M. (2007). Compressed sensing MRI. \textit{IEEE Signal Processing Magazine}, 25(2), 72-82.
\bibitem{ref40}
M. Uecker et al., ``ESPIRiT—an eigenvalue approach to autocalibrating parallel MRI: where SENSE meets GRAPPA,'' \textit{Magn. Reson. Med.}, vol. 71, no. 3, pp. 990--1001, May 2014.
\bibitem{ref41}
R. M. Henkelman, ``Measurement of signal intensities in the presence of noise in MR images,'' \textit{Med. Phys.}, vol. 12, no. 2, pp. 232--233, Mar. 1985.
\bibitem{ref42}
C. E. Hayes and P. B. Roemer, ``Noise correlations in data simultaneously acquired from multiple surface coil arrays,'' \textit{Magn. Reson. Med.}, vol. 16, no. 2, pp. 181--191, Nov. 1990.
\bibitem{ref43}
A. C. Aitken, ``IV.—On least squares and linear combination of observations,'' in \textit{Proc. R. Soc. Edinb.}, vol. 55, pp. 42--48, Sep. 1936.
\bibitem{ref44}
L. Chen, X. Chu, X. Zhang, and J. Sun, ``Simple baselines for image restoration,'' in\textit{European conference on computer vision}, pp. 17--33, Oct. 2022.
\bibitem{ref45}
X. Glorot and Y. Bengio, ``Understanding the difficulty of training deep feedforward neural networks,'' in \textit{Proc. Conf. Artificial Intelligence and Statistics}, vol. 9, pp. 249--256, May 2010.
\bibitem{ref46}
D. P. Kingma and J. Ba, ``Adam: A method for stochastic optimization,'' in \textit{International Conference on Learning Representations}, May 2015.
\bibitem{ref47}
A. Paszke et al., ``Pytorch: An imperative style, high-performance deep learning library,'' in \textit{Proc. Adv. Neural Inf. Process. Syst.}, vol. 32, pp. 8026--8037, Dec. 2019.
\bibitem{ref48}
J. Zbontar et al., ``fastMRI: An open dataset and benchmarks for accelerated MRI,'' \textit{arXiv preprint}, arXiv:1811.08839, Nov. 2018.
\bibitem{ref49}
S. M. Smith, ``Fast robust automated brain extraction,'' \textit{Hum. brain map.}, vol. 17, no. 3, pp. 143--155, Sep. 2002.
\bibitem{ref50}
S. Sengupta et al., “A specialized multi-transmit head coil for high resolution fMRI of the human visual cortex at 7 T,” \textit{PLoS ONE}, vol. 11, no. 12, Dec. 2016, Art. no. e0165418.
\bibitem{ref51}
T. W. Redpath, ``Noise correlation in multicoil receiver systems,'' \textit{Magn. Reson. Med.}, vol. 24, no. 1, pp. 85--89, Mar. 1992.
\bibitem{ref52}
P. B. Roemer, W. A. Edelstein, C. E. Hayes, S. P. Souza, and O. M. Mueller, ``The NMR phased array,'' \textit{Magn. Reson. Med.}, vol. 16, no. 2, pp. 192--225, Nov. 1990.

\end{thebibliography}
\end{document}